# Dissolved gas monitoring probe without liquid-gas separation under strong electromagnetic interference

## Author Information

### Affiliations

**State Key Laboratory of Alternate Electrical Power System with Renewable Energy Sources, North China Electric Power University, Beijing, P.R. China**

Guo-ming MA, Yuan WANG, Zhang-lin CHEN, Yang-yang XIE, Wei-qi QIN & Diya ZHENG

**Guangdong Engineering Technology Research Centre of Power Equipment Reliability in Complicated Coastal Environments, Tsinghua Shenzhen International Graduate School, Tsinghua University, Shenzhen, Guangdong, P.R. China**

Hong-yang ZHOU

**College of Engineering, Peking University, Beijing, China**

Hao ZHAO

**Department of Mechanical and Aerospace Engineering, Princeton University, Princeton, New Jersey, US**

Chao YAN

### Contributions

G.M. proposed and coordinated the work. Y.W. jointly developed the idea and simulation models, design gas system and carried out experiments. Z.C. and Y.X. modified the simulation models, carried out experiments. Y.W., W.Q. and H.Z. proposed the phase demodulation algorithm and performed data analysis. G.M., Y.W. and D.Z. completed microchannel processing and fusion splicing. H.Z. and C.Y. modified the idea and the set-up.

### Corresponding author

Correspondence to: Guo-ming MA

## Abstract

Rapid and direct monitoring of dissolved gases in liquids under strong electromagnetic interference is very important. Electronic gas sensors that can be placed into liquids are difficult to work reliably under strong electromagnetic fields. The existing optical monitoring techniques for dissolved gases all require gas-liquid separation, and are conducted in gas phase, which have poor timeliness. In this paper, a dissolved gas monitoring probe without liquid-gas separation under strong electromagnetic interference is proposed. We take transformer oil-dissolved acetylene monitoring as an example, an oil-core photonic crystal fiber photothermal interferometry probe is proposed and demonstrates the feasibility of trace oil-dissolved acetylene directly monitoring without oil-gas separation. The minimum detection limit reaches 1.4 ppm, and the response time is about 70 minutes. Due to the good insulation performance and the compact size, the all-fiber probe provides applicability to be placed into transformer oil and perform on-line monitoring under strong electromagnetic interference.

## Introduction

Rapid monitoring of dissolved gases in liquids has important implications for biomedical, aerospace, and energy fields. In order to realize the rapid detection of dissolved gas, researchers hope to build the sensor into the liquid for direct detection. However, electronic sensors are limited to being susceptible to strong electromagnetic interference (EMI), and it is difficult to work reliably in a strong electric field environment. Therefore, how to realize the rapid and accurate monitoring of dissolved gas in liquid under strong EMI has important research value. This paper takes the monitoring of dissolved gas in power transformer oil as an example to illustrate the novel monitoring probe.

Power transformers often work in strong electromagnetic field environments[1]. If there is a fault inside the power transformer, the transformer oil will accelerate the deterioration and decompose to generate small molecular hydrocarbon gases[2]. The new generated gases will continue to dissolve in the oil through convection and diffusion. Since acetylene is one of the most important characteristic gases, the monitoring of the dissolved acetylene in transformer oil allows the evaluation of transformer operating conditions[3].

Nowadays, gas chromatography (GC) is a widely applied technique for dissolved gas analysis (DGA). However, GC is an off-line test method with low detection frequency, which makes it difficult to detect transformer faults in time. Also, GC has some other demerits, such as the overall operation process is complicated, the detection results are widely dispersed, and the chromatographic columns are prone to contamination. So, it is not suitable for online monitoring. Recently, optical gas sensors are gradually get used in-situ online monitoring for its strong anti-electromagnetic interference ability[4]. Laser absorption spectroscopy is a commonly used technique which is mainly based on Beer-Lambert law. It distinguishes gases by the 'finger-print' absorption lines of different gas molecules, and the minimum detectable limit (MDL) is greatly affected by the absorption pathlength. For a better performance in MDL, researchers introduced multi-pass gas cells into gas detection. For example, J.M. Dai et al developed a tunable diode laser absorption spectroscopy (TDLAS) gas sensing setup with a White cell[5], which achieved an accurate measurement of 1 ppm acetylene. Similarly, G.M. Ma et al used a Herriott gas cell into trace acetylene sensing[6]. Although multi-pass gas cells increase the absorption pathlength effectively, they are usually bulky and require optical calibration.

Hollow-core photonic crystal fibers (HC-PCFs) are microstructed fibers[7-9]. Nowadays, HC-PCF is popular in gas sensing due to its core can act as a gas cell, and it can be coiled to a small size with low dispersion and sufficient light-gas interaction over long distances[10-14]. W. Jin et al designed a fiber optic gas detection system for transformer monitoring with a 1 m HC-PCF was used as a gas cell[15]. The system measured the spectral attenuation and achieved an MDL of 1 ppm for acetylene. In order to further improved the sensitivity of the sensing system, phase change induced by photothermal effect in HC-PCF has been exploited in gas sensing[16-19]. When laser gets absorbed by gas molecules, there will be a local temperature rise, and the refractive index will change accordingly[4,19-22]. Photothermal interferometer (PTI) measures the accumulated phase change of light over propagation distance and has proven to be very sensitive in trace gas sensing. W. Jin et al firstly introduced an all-fiber PTI system with a 0.62 m HC-PCF gas cell, and demonstrated a MDL of 30 ppb for acetylene[17]. The sensitivity of PTI system is much better than the other techniques.

However, when gas detection techniques get applied to DGA, they all need to be completed in the gas phase, and it is impossible to directly monitoring acetylene in transformer oil. It should be noted that due to the large oil tank, it takes a long time for oil-dissolved acetylene to diffuse to the sampling port. This process may take tens of hours or even days, making it difficult to measure the gas in time. Also, the result is greatly affected by the sampling and degassing process, and the analysis results may have large errors. Therefore, it is of great significance to carry out the detection of oil-dissolved acetylene directly in power transformer.

In this paper, we report an oil-core photonic crystal fiber (OC-PCF) photothermal interferometry probe to monitoring oil-dissolved acetylene. The OC-PCF confines the fluidic sample and propagating light within its oil-core. The proposed probe achieves fast response and high sensitivity to oil-dissolved acetylene and is immune to the external noise.

## Results

### Analysis of the OC-PCF

In this paper, the OC-PCF was fabricated from the NKT Photonics' HC-1550-02 fiber with a core diameter of about 11 μm and a length of about 0.85 m. Figures 1a and 1b show the structure of an OC-PCF and the scanning electron microscope image of its cross-section. There are 4 microchannels distributed evenly along its axial direction, and the outer size of the microchannels were about 3 μm×3 μm (see Methods). In addition, both ends of PCF are fusion spliced with single mode fibers (SMFs), and the medium can only flow into the core region through microchannels. When the PCF is placed in the transformer oil (RI=1.4745), the transformer oil can only enter the core region and the cladding air holes penetrated by microchannels, and an OC-PCF is made. The spatial light field intensity distribution in an OC-PCF is shown in Fig. 1d. At this time, the effective mode refractive index of an OC-PCF is about 1.4714.

**Fig. 1 Principle of PTI in OC-PCFs**

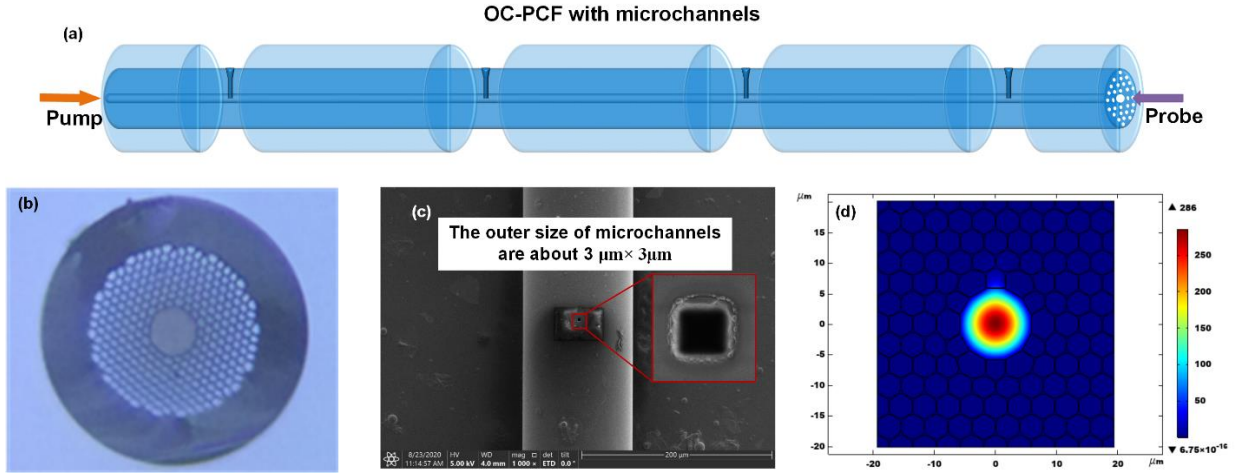

**a**, Conceptual view of an OC-PCF with microchannels along its axial direction. **b**, Scanning electron microscope image of the NKT Photonics' HC-1550-02 fiber used in this work. **c**. Scanning electron microscope image of a microchannel on a PCF. **d**, The spatial optical mode intensity distribution in an OC-PCF.

Since the effective refractive index of the cladding is lower than that of the oil core, the oil-core photonic crystal fiber can be equivalent to a traditional step-index fiber, so its light guiding mechanism can be explained by the total internal reflection mechanism[23-25]. The light guiding performance of OC-PCF is enhanced compared to HC-PCF.

From the above simulation results, it can be seen that when part of the cladding air holes and the hollow core of a HC-PCF are filled with transformer oil, the two-dimensional photonic crystal structure of the cladding quartz-air holes remains basically unchanged, and the optical energy is basically distributed in the central oil core. When the laser is transmitted in the oil core, it can fully interact with the oil sample.

## Theory of PT spectroscopy in OC-PCF

The basic schematic diagram in an OC-PCF can be shown in Fig. 1a. When an intensity modulated pump light travels in the OC-PCF, it interacts with the oil sample and get absorbed by the oil-dissolved acetylene. According to the photothermal effect, a periodic heat source will be generated, which modulates local temperature and RI of the oil sample. A constant probe light counter propagates with the modulated pump light and measures the phase change induced by the periodic heat source.

Usually, the fundamental mode profile of the light in OC-PBF follows Gaussian distribution, so it is assumed that the intensity profile of the pump light can be expressed as

$$I(r,t) = \frac{2P_{pump}}{\pi w_{pump}^2} \exp(-2r^2/w_{pump}^2) \cdot S(t) \tag{1}$$

Where, $P_{pump}$ is the peak power of the pump light, $w_{pump}$ is the mode field radius, $S(t)$ is the intensity modulation function. In this paper, $w_{pump}$ is set to 4.03 m, and $S(t)=1+\sin(2\pi f_m t)$, $f_m$ is the modulated frequency. When the pump light interact with the oil-dissolved acetylene, each acetylene molecule is equivalent to a heat source. Therefore, the periodic heat source may be written into

$$Q(r,t) = \alpha_0 C I(r,t) \tag{2}$$

Where, $\alpha_0$ is the absorption coefficient of acetylene, $C$ is the acetylene concentration.

In order to study the temperature rise in the oil-core caused by the photothermal effect, the following assumptions are made in this paper. Since acetylene is evenly distributed in the OC-PCF, the thermodynamic process in the OC-PCF can be regarded as consistent with the thermodynamic process in the continuum. The light absorption of trace acetylene gas is very weak. It may be considered that the light intensity of the pump light is always constant along the light propagation direction. The light intensity near the boundary between the oil-core and the cladding is very small, so the temperature at the boundary remains constant at the ambient temperature.

According to the above assumption, the temperature distribution within the oil-core can be obtained by solving the heat transfer equation

$$\frac{\partial^2 \Delta T(r,t)}{\partial r^2}+\frac{1}{r}\frac{\partial \Delta T(r,t)}{\partial r}+\frac{1}{\kappa}Q(r,t)=\frac{\rho C_p}{\kappa}\frac{\partial \Delta T(r,t)}{\partial r},(t>0,0\leq r<b) \qquad (3)$$

Where, $\kappa$ is the thermal conductivity of transformer oil, the value is 0.128 W/(m·K); $\rho$ is the the density of transformer oil, equals to 895 kg/m³ at room temperature; $C_p$ is the specific heat capacity of the transformer oil, the value is 2100 J/(kg·K); $b$ is the radius of the oil-core.

When the acetylene concentration is 100 ppm, the absorption coefficient is 1.16 cm$^{-1}$, the pump light power is 40 mW, and the modulation frequency is 20 kHz, the relationship between the temperature change at the center of the oil-core and the pump light interaction time is shown in Fig. 2a. It can be seen that, with the increase of the interaction time, the temperature in the oil-core gradually increased, and the temperature became stable after about 10 ms. Since the intensity modulation frequency of the pump light is 20 kHz, the temperature also changes sinusoidally with a period of 20 kHz. After the temperature reaches a steady state, Fig. 2b shows the temperature rise distribution within the oil-core, and Fig. 2c shows the temperature distribution along the OC-PCF. We can know that the maximum temperature rise occurs in the center of the oil core, and the maximum temperature rise is about 2.4 K. Also. the temperature of each point along the fiber is basically the same.

**Fig. 2 Computation results of temperature distribution in OC-PCF.**

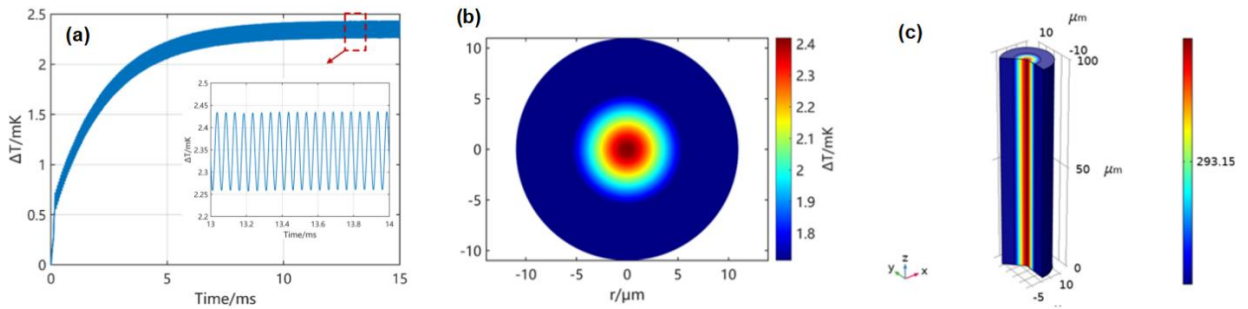

a. Temperature rising at the center of the OC-PCF. b. Temperature rising distribution within the oil-core. c. Temperature distribution along the OC-PCF.

When the local temperature changes, the refractive index of the transformer oil will change accordingly, which satisfies[26]

$$\Delta n = \Delta T \cdot \partial n / \partial T \tag{4}$$

Therefore, the phase change of the probe light due to the photothermal effect can be obtained by[4]

$$\Delta\varphi(t) = \frac{2\pi l}{\lambda_{probe}} \cdot \Delta n_{eff} = \frac{4l}{w_{probe}^2 \lambda_{probe}} \frac{\partial n}{\partial T} \int_0^b \Delta T(r,t) \cdot \exp\left(-2r^2/w_{probe}^2\right) \cdot 2\pi r dr \tag{5}$$

Where, $l$ is the optical pathlength of probe light in OC-PCF, $\lambda_{probe}$ is the wavelength of the probe light, $w_{probe}$ is the mode field radius, $\Delta n_{eff}$ is the effective refractive index.

In summary, when the pump light interacts on the oil sample with acetylene dissolved in it, the photothermal effect can still be effectively excited. Therefore, the phase modulation changes accumulated by the probe light in the OC-PCF can be used to reflect the concentration of the oil-dissolved acetylene.

## Experimental setup

Fig. 3 depicts the experimental setup of the PTI acetylene monitoring based on OC-PCF. A distributed feedback laser (DFB) is used as a pump laser. It is modulated at 20 kHz. A RIO laser is as a probe laser. The OC-PBF forms the sensing arm of a Mach-Zehnder interferometer (MZI) while the reference arm is made of a single-mode fiber (SMF). The probe light is split into two beams and travels in the sensing arm and the reference arm. Two beams of the probe light will be recombined by a fiber coupler and interfere with each other at a balanced photodetector.

For limiting the 1/$f$ noise of the PTI probe, the optical frequency in reference arm is up-shifted by 200 MHz with the help of an acousto-optic modulator (AOM)[27]. A heterodyne beatnote at 200 MHz will be collected by DAQ, and the modulated phase accumulation will be accurately demodulated through the differential cross-multiplication (DCM) algorithm (see Supplementary Discussion) and fast Fourier transform (FFT).

According to Eqn. (5), for the purpose of improving the sensitivity of the PTI probe, one possible solution is to increase the optical pathlength of the probe light in the OC-PCF. Therefore, a fiber loop is introduced in the sensing arm through the fiber couplers. The pump light is coupled into the fiber loop through a optical circulator (OC) and gets stopped by that same OC after one revolution. In contrast, the probe light freely travels in the fiber loop, and the pathlength in the OC-PCF is greatly increased.

A fiber Bragg grating (FBG), whose central wavelength equals to the wavelength of the probe light, is introduced into the sensing arm. It is used as an optical filter. Since the central wavelength of the FBG equals to the wavelength of the probe light, the probe light will be reflected by the FBG and enter the sensing arm again, while the residual pump light cannot be reflected by the FBG and will be filtered out.

**Fig. 3 Experimental setup of PTI acetylene sensing with 0.85-m-long OC-PBF. DFB, distributed feedback laser, used as pump source; EDFA, erbium-doped fiber amplifier; AWG, arbitrary waveform generator; FC1-FC4, fiber couplers, and the splitting ratios are all 50/50; OC, optical circulator; FBG, fiber Bragg grating, center wavelength is 1550 nm; AOM, acousto-optic modulator; BPD, balanced photodetector; DAQ, data acquisition.**

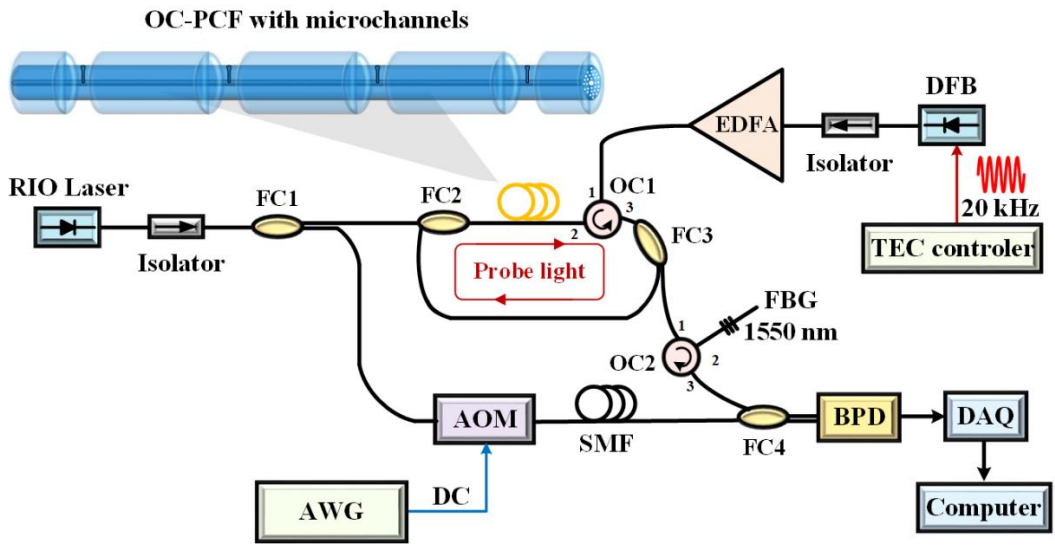

Test of minimum detection limit

To stimulate the photothermal effect, the wavelength of the pump light is tuned to 1530.37 nm. The absorption coefficient at this wavelength is 1.16 cm$^{-1}$ at room temperature and 1 atm[28]. The wavelength of
99

the probe light is set to 1550 nm, where the absorption coefficient is much smaller than that at 1530.37 nm. In order to amplify the optical power of the pump light, the working mode of the EDFA is set to AGC mode, and the amplification factor is constant at 21.0 dB. Thus, the pump light power coupled into the sensing arm is about 16 dBm.

However, the operating environment of the substation is complex, and there are many external factors such as temperature changes and vibrations. In addition, when a fault such as partial discharge occurs in the power equipment, it is often accompanied by the generation of ultrasonic waves[29,30]. When the ultrasonic signal acts on the sensing arm of the MZI, it will also cause the phase of the probe light get modulated[31]. Since the modulation frequency of the pump light in this paper is set as 20 kHz, and the frequency of ultrasonic wave in power equipment is mostly range from 20 kHz to 200 kHz[32,33], the ultrasonic signal generated by the faults of the power equipment will affect the monitoring accuracy of the OC-PCF photothermal interferometry probe. What is more, the irradiation of pump light with high optical power in the transformer oil may also introduce temperature rise caused by non-photothermal effect, making the monitoring result inaccurate.

To solve the above problems, the dual-pump-wavelength differential method is introduced in gas monitoring[34-37]. In this method, the DFB is used to output two different pump light with wavelengths $\lambda_1$ and $\lambda_2$ by changing the temperature. Gas detections with two different wavelengths conduct at almost the same time. The phase change caused only by the photothermal effect can be obtained by taking the difference of the phase demodulation results at the two wavelengths. The wavelength $\lambda_1$ is located at the peak of the acetylene absorption line, and the wavelength $\lambda_2$ is located at the weakest point of the acetylene absorption line, and $\lambda_2$ is almost equals to $\lambda_1$. Through the dual-pump-wavelength differential method, the influence of external interference and photoelectric device drift on the monitoring results can be effectively reduced.

According to the HITRAN database, in this paper, $\lambda_1$ is set to 1530.37 nm, as sensing pump wavelength; and $\lambda_2$ is set to 1530.71 nm, as reference pump wavelength[28]. Multiple PTI tests were performed on oil samples with an acetylene concentration of 400 ppm (see Methods), and an ultrasonic signal source was placed near the oil cell. The average phase demodulation result of the sensing pump light at 20 kHz was about 3.632 mrad; the average phase demodulation result of the reference pump light was about 0.409 mrad, and the phase difference between the two wavelengths was about 3.223 mrad. When oil-dissolved acetylene was detected by sensing pump light without any external ultrasonic waves, and the average result at 20 kHz was about 3.311 mrad.

**Fig. 4: Results based on dual-pump-wavelength differential method**

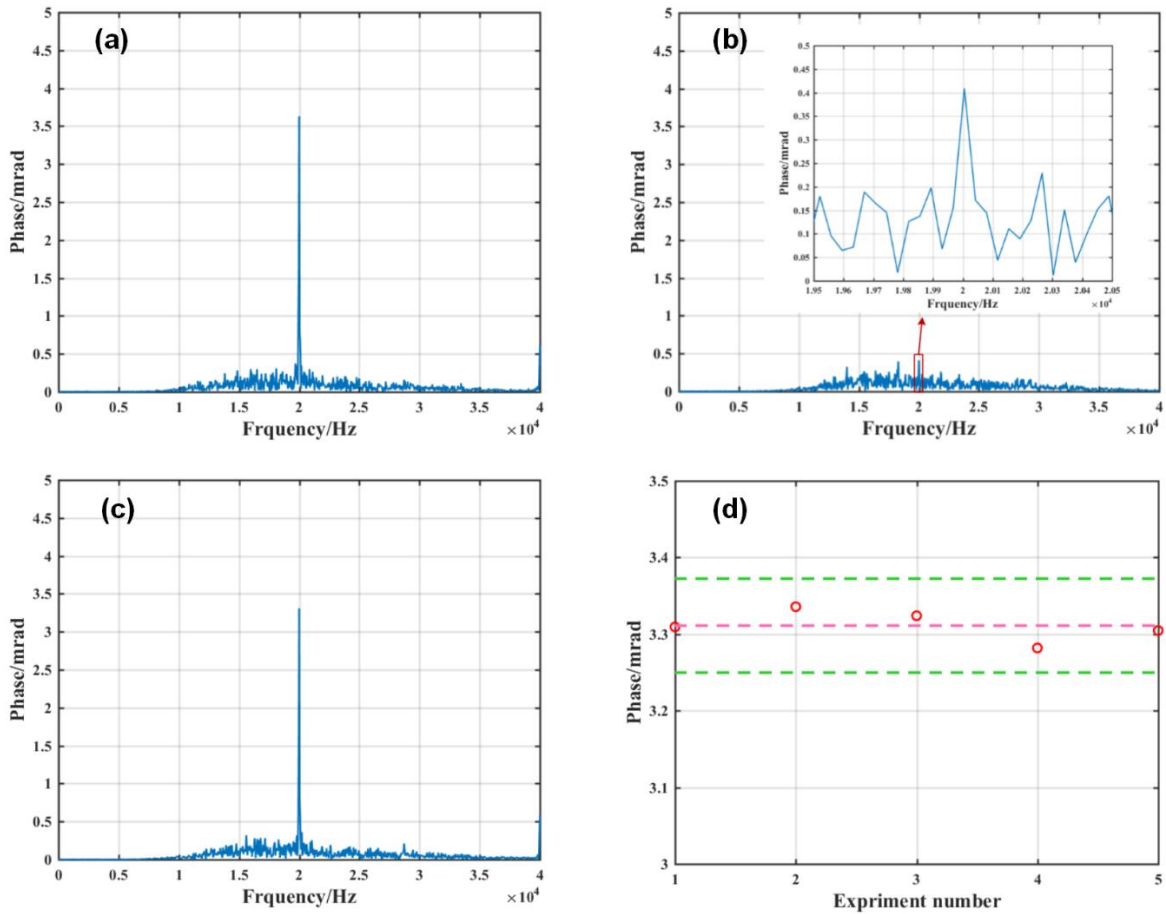

**a.** The demodulation result when the wavelength of the pump light is 1530.37 nm (the external ultrasonic frequency is 20kHz). **b.** Demodulation result when the wavelength of pump light is 1530.71 nm (the external ultrasonic frequency is 20kHz). **c.** Demodulation result when the wavelength of pump light is 1530.37 nm without external noise. **d.** Monitor results for multiple times

It can be seen that the phase change caused only by the photothermal effect can be accurately obtained by dual-pump-wavelength differential method, which significantly reduces the serious influence of external noise on the accuracy of acetylene monitoring.

In order to test the MDL of the OC-PCF photothermal interferometry probe based on the dual-pump-wavelength differential method, the OC-PCF was first placed into an oil cell filled with 1500 mL pure transformer oil. And then $C_2H_2/N_2$ gas mixtures with different concentration were injected into the transformer oil. The oil-dissolved acetylene detections were carried out at room temperature.

The results show that under the same concentration of oil-dissolved acetylene, the demodulated phase difference at 20 kHz obtained by the dual-pump-wavelength differential method has small fluctuation and good stability. In addition, as the oil-dissolved acetylene concentration increases, the demodulated phase difference at 20 kHz increases accordingly. The monitoring results under different oil-dissolved acetylene concentrations is shown in Fig. 5

**Fig. 5: The monitoring results under different oil-dissolved acetylene concentrations with a 0.62-m-long OC-PCF.**

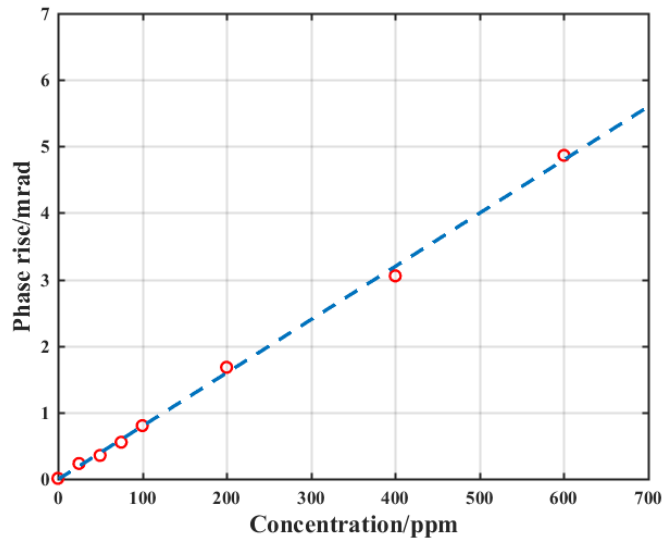

There is a good linear relationship between demodulated phase differences under dual pump wavelengths and oil-dissolved acetylene concentrations, and the linear correlation coefficient R-square is 0.992. The fitted equation can be given as:

$$\Delta\varphi = 8.16\times10^{-3} C(\text{C}_2\text{H}_2) - 0.001 \qquad (6)$$

Thus, the sensitivity of the proposed probe is about $8.16\times10^{-3}$ mrad/ppm. The oil-dissolved acetylene concentration can be back-calculated using the monitoring results based on the Eqn. (6). When the oil-dissolved acetylene concentration was 50 ppm, the demodulated phase difference was about 0.394 mrad, while the standard deviation is about 0.011 mrad. The MDL in terms of 1σ noise equivalent concentration[38] can be estimated to be 1.396 ppm.

## Test of multi-gases cross-sensitivity

During the decomposition of transformer oil, some other characteristic gases such as methane, ethylene, carbon monoxide, and carbon dioxide are also generated. The other characteristic gases present in the transformer oil may also be excited by the pump light to produce photothermal effects, which affect the detection accuracy. Therefore, it is necessary to test the cross-sensitivity of other characteristic gases to the proposed OC-PCF photothermal interferometry probe.

A specific concentration of gas mixture was injected into the transformer oil. The main gas components are hydrogen (601 ppm), methane (120 ppm), ethylene (113 ppm), ethane (118 ppm), acetylene (62.6 ppm), carbon monoxide (611 ppm), and carbon dioxide (2863 ppm). The oil-dissolved acetylene was monitored by the proposed OC-PCF photothermal interferometry probe and GC (Agilent 7890B). The results at dual wavelength are shown in Fig. 6. The phase difference under the dual wavelength was about 0.419 mrad, and the oil-dissolve acetylene concentration in the above oil sample can be back-calculated from Eqn. (6) to be about 55.012 ppm.

The oil samples were detected twice by gas chromatograph, and the acetylene concentrations were 58.171 ppm and 53.198 ppm, which were basically the same as the acetylene detection results obtained by the proposed OC-PCF photothermal interferometry probe. The maximum error did not exceed ±4 ppm, which can meet the accuracy requirements for DGA online monitoring in IEEE Std C57.104-2019[2].

At the same time, when the pump light wavelengths were selected as 1530.37 nm and 1530.71 nm, the other characteristic gases did not have any adversely influences on the monitoring results of oil-dissolved acetylene. The above results verify that the proposed OC-PCF photothermal interferometry probe has excellent detection accuracy and does not have the problem of multi-gases cross-sensitivity.

**Fig. 6 Results of multi-gases cross-sensitivity test**

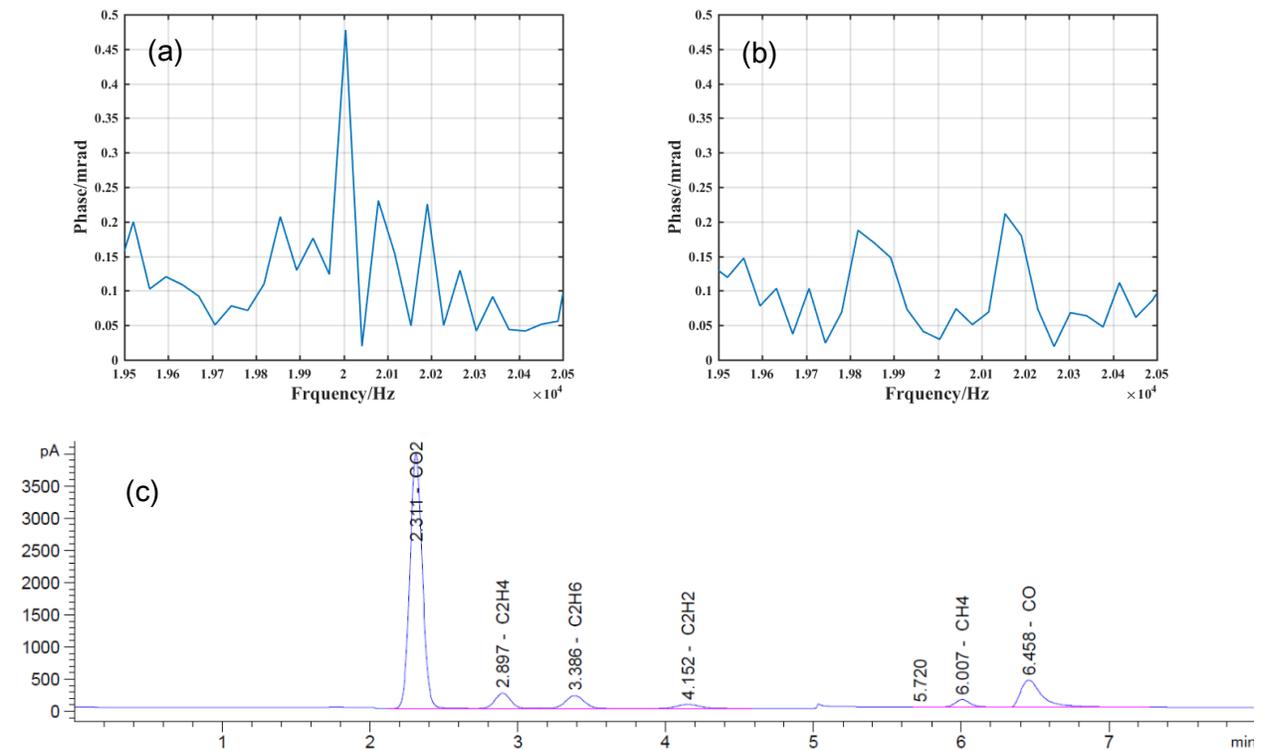

**a.** Demodulation result when the wavelength of pump light is 1530.37 nm. **b.** Demodulation result when the wavelength of pump light is 1530.71 nm. **c.** Result of gas chromatography

## Test of response time

In order to study the time response of the OC-PCF photothermal interferometry probe, a $C_2H_2/N_2$ gas mixture at a concentration of 75 ppm was continuously injected into the oil sample with an initial acetylene concentration of 50 ppm for 180 min, and then a 50 ppm $C_2H_2/N_2$ gas mixture was continuously injected for 180 min. During this period, oil-dissolved acetylene was detected by the proposed OC-PCF probe every 10 min, and the phase demodulation results at 20 kHz under dual wavelengths were recorded. The specific results are shown in Fig. 7.

The demodulated phase amplitude begins to change after about 30 min since the gas mixture got injected into the transformer oil. This is because the dissolution and diffusion of acetylene is not completed instantaneously, and it can be considered that the acetylene concentration in the oil at the OC-PCF changes after 30 min. After that, the phase changes gradually with the increase of gas injection time until it reaches stability.

**Fig. 7 Results of time response test**

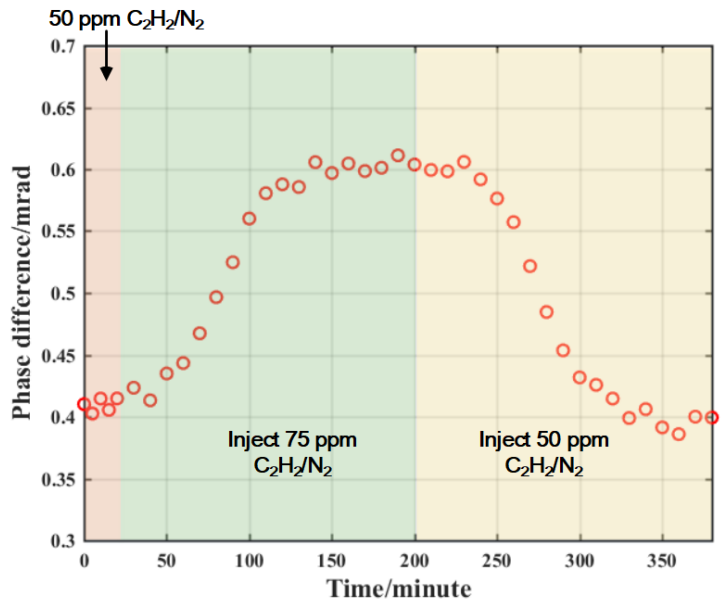

Defining the time taken to reach 90% of the maximum phase change from 10% of the maximum phase change as the response time of the OC-PCF photothermal interferometry probe. Therefore, the response time of the OC-PCF photothermal interferometry probe is about 70 min.

Since the proposed probe can be built into the transformer, it is not necessary to take samples at the oil outlet and conduct oil-gas separation, which greatly improves the real-time performance of the monitoring system and greatly shortens the fault discovery period to 70 min.

## Discussion

To solve the problems that the existing dissolved gas analysis techniques are difficult to work under strong electromagnetic interference and over-reliance on liquid-gas separation, a dissolved gas monitoring fiber-based probe is proposed. This paper takes the monitoring of transformer oil-dissolved acetylene as an example, an online monitoring probe based on oil-core photonic crystal fiber and photothermal interferometry technique is proposed for the first time. This scheme achieves to directly detect oil-dissolved acetylene in the oil-core of an OC-PCF, there is no need to get acetylene separated from transformer oil. The OC-PCF is made of NKT Photonics' HC-1550-02 fiber and transformer oil. In this paper, we successfully demonstrate that laser can travel in the oil-core efficiently and can fully interact with the oil sample. This is because the transformer oil only enters the PCF core and a small amount of air holes in the cladding, there is no significant change in the dielectric coefficient distribution of the cladding, so light can be transmitted in the OC-PCF based on the total internal reflection light guiding mechanism. What is more, based on the thermodynamic model, we find that when the pump light interacts on the oil sample with acetylene dissolved in it, the photothermal effect can still be effectively excited. These provide the possibility to monitor oil-dissolved acetylene without oil-gas separation.

To achieve high sensitivity and low detection limit for direct oil-dissolved acetylene monitoring, this paper proposes to introduce a fiber loop into the sensing arm of the Mach-Zehnder interferometer, which effectively increases the effective optical pathlength of the probe light in OC-PCF. And we use dual-pump-wavelength differential method to eliminate the adversely influence caused by external noise (such as vibrations in transformer, ultrasonic signal caused by partial discharge and frequency noise introduced by probe laser) on the monitoring results. The detection of the oil-dissolved acetylene on the order of µL/L can be realized. Experiments have proved that the response time of the proposed OC-PCF photothermal interferometry probe is about 70 min, the minimum detection limit of acetylene reaches 1.39 ppm, and the measurement error at low concentration does not exceed ±4 ppm. The results verifies that the developed OC-PCF photothermal interferometry probe has high real-time performance, high sensitivity, low dispersion, and can meet the needs of oil-dissolved acetylene monitoring in actual power transformer.

The OC-PCF-based photothermal interferometry probe can be easily applied to monitor other characteristic gases in transformer oil. Since the transmission window of HC-1550-02 ranges from 1490 nm to 1680nm, it covers a range of gases absorption lines of such as $CO_2$, CO, $CH_4$, etc. Therefore, the direct detection of multiple gases can be realized only by reasonably changing the wavelength of the pump light. It is very beneficial to the research and development of all-component transformer oil dissolved gases on-line monitoring system.

In addition, the interior of the central core region of the photonic crystal can also be filled with different liquids according to actual needs, so as to meet the direct measurement of dissolved gases in different application environments.

## Methods

### Preparation of OC-PCF

Both ends of the HC-PCF (NKT Photonics' HC-1550-02 fiber) were fusion spliced with traditional SMFs (Corning' SMF-28) with the Fujikua LZM-100 fusion splicer. After splicing, the optical transmission loss at the wavelength of 1550 nm was only about 3.1 dB. Four microchannels were drilled along a 0.85-m-long HC-PCF by focused ion beam (FIB) technique and were spaced 21 cm from each other (see Supplementary Fig. 1).

In this paper, a Strata 400S FIB (Thermo Fisher Scientific) with $Ga^+$ ion source was applied. Due to the large aspect ratio of microchannels, it is difficult to achieve low-loss, small-diameter microchannel processing. Therefore, the acrylic polyester coating layer on the surface of HC-PCF should be first removed before FIB processing. The outer size of the microchannels is about 3 μm×3 μm. The loss introduced by a single microchannel is only about 0.13 dB at the wavelength of 1550 nm.

Place the processed HC-PCF into transformer oil, and the transformer oil will flow into the hollow core region through microchannels, and finally an OC-PCF gets formed.

## The waveguide simulations

Simulations of the optical waveguide was conducted by using COMSOL Multiphysics 2D 'Electromagnetic waves, Frequency Domain' module. The silica refractive index $n_{SiO_2} = 1.45$, the air refractive index $n_{air} = 1$ and the transformer oil refractive index $n_{oil} = 1.4745$ were entered into the calculation. The effective mode refractive index of OC-PCF was calculated about 1.4714 at a wavelength of 1.55 μm.

## Preparation of oil samples

At room temperature and atmospheric pressure, by using two mass flow controllers to control the flow speed of acetylene and nitrogen, the preparation of gas mixture with different volume concentrations is achieved. The mixed gas is then continuously injected into the transformer oil by an aeration tube and stirred by a mechanical stirrer to accelerate the dissolution.

## **Data Availability**

All data in main text and supplementary information is available from the corresponding authors upon reasonable request.

## **Code Availability**

The codes that were used in this study are available upon request to the corresponding author.

## Acknowledgements

This work was supported by National Natural Science Foundation of China (51677070, 51977075), Young

Elite Scientists Sponsorship Program by CAST YESS20160004, Fok Ying-Tong Education Foundation for

## Ethics declarations

The authors declare no conflict of interest.